Strain-induced energy band gap opening in two-dimensional bilayered silicon film

Z. Ji[1], R. Zhou[2], L. C. Lew Yan Voon[3], Y. Zhuang[1]

[1]Department of Electrical Engineering, Wright State University, Dayton, OH 45435, USA

[2]Department of Electrical and Computer Engineering, Purdue University, West Lafayette, Indiana 47907, USA

[3]School of Science and Mathematics, The Citadel, Charleston, South Carolina 29409, USA

Abstract

This work presents a theoretical study of the structural and electronic properties of bilayered silicon films under in-plane biaxial strain/stress using density functional theory. Atomic structures of the two-dimensional silicon films are optimized by using both the local-density approximation and generalized gradient approximation. In the absence of strain/stress, five buckled hexagonal honeycomb structures of the bilayered silicon film have been obtained as local energy minima and their structural stability has been verified. These structures present a Dirac-cone shaped energy band diagram with zero energy band gaps. Applying tensile biaxial strain leads to a reduction of the buckling height. Atomically flat structures with zero bucking height have been observed when the AA-stacking structures are under a critical biaxial strain. Increase of the strain between 10.7% ~ 15.4% results in a band-gap opening with a maximum energy band gap opening of ~168.0 meV obtained when 14.3% strain is applied. Energy band diagram, electron transmission efficiency, and the charge transport property are calculated.

**Introduction**

The appearance of graphene has attracted tremendous interests and has found particular applications in ultrahigh-speed electronics due to its massless Dirac carriers and high electron mobility[1-4]. A graphene-based field effect transistor with a record high cutoff frequency up to 300GHz has been reported[5]. The major bottleneck for graphene is the lack of an energy band gap. Other forms of two dimensional (2D) materials such as $MoS_2$[6-7], germanene[8-10], borophene[11], and phosphorene[12-15] have been increasingly investigated theoretically and experimentally. Silicene, a single atomic layer of silicon with a buckled hexagonal aromatic stage if free standing, shows the greatest potential given to its natural compatibility with the silicon technology[16]. Similar to its carbon counterpart graphene, the Dirac cone shaped energy band diagram and consequently the zero energy band gap impedes silicene in the main stream semiconductor industry. Although a band gap opening in silicene has been reported by applying an external vertical electrical field[8, 17-18], the required electrical field exceeds the dielectric strength of most of the dielectric materials except diamond, which makes this method impractical.

On the other hand, strain/stress has been widely used as a means to improve electron and hole mobilities in silicon (Si) technology[19-22]. A few works have addressed the impact of strain/stress on the energy band diagram of 2D materials[23-29]. Band gaps up to 1.62 meV and 1.94 meV have been reported in [27] by applying uniaxial tensile stress along the armchair and zigzag direction, respectively. However, such small band gap openings impose a formidable challenge in making thermally stable devices at room temperature. Biaxial strain/stress does not induce any band gap opening in silicene[28].

In this work, we report calculations of the structural and electronic properties of bilayered silicon film (BiSF) under biaxial stress. Five energetically favorable atomic structures of free-standing BiSF have been obtained. In the absence of strain/stress, the BiSF shows Dirac-cone shaped energy band diagrams with zero energy band gaps. Applying a tensile strain between 10.7% ~ 15.4% leads to a band-gap opening with a maximum of ~168.0 meV.

**Method**

The DFT calculations were performed using Atomistix Tool-Kit (ATK, QuantumWise). The atomic structure has been optimized using both the local density approximation (LDA) and the generalized gradient approximation (GGA). In order to minimize the impact from the adjacent Si layers, the lattice constant along the out of plane i.e. z-axis was set to 25 Å. A 21×21×1 Monkhorst-Pack grid k-point mesh has been used along the x-, y-, and z- axis for the Brillouin zone sampling. In all the calculations, the force and stress tolerance, density mesh cut-off and electron temperature were set at the value of 0.0001 eV/Å, 75 Hartree and 300K, respectively. Also the non-polarized correlation was employed for the calculation. Periodic boundary condition has been employed for unit cells 1×1, 2×2, and 3×3 when optimizing the atomic structure. The transport properties were calculated by using FFT in solving the non-equilibrium Green's function (NEGF) with norm-conserving pseudo-potentials.

**Results and Discussions**

A BiSF unit cell contains four silicon atoms and is considered as a free standing structure during the simulations. Structural optimization has been performed by taking the four silicon atoms freely relaxed along the x-, y-, z- axis. The Si-Si bond length is 2.342 Å and the lattice constant is 3.866 Å. Its validity is confirmed by the simulation results obtained using DFT- GGA and LDA approximations are in very good agreement. Five energetically stable structures with AA-stacking (Fig. 1(a)) and AB-stacking (Fig. 1(b)) are obtained and their stability has been verified: i) all silicon atoms in the five structures are found to return to their optimized positions after applying a random perturbation; ii) in addition to the 1×1 unit cell, structural optimization calculations have also been performed on 2×2, and 3×3 supercells to further verify the stability. The four silicon atoms with atoms A1 and A2 in the upper layer, and A3 and A4 in the lower layer are coplanar in AA-stacking. Depending on the orientation of the A1-A2 and A3-A4 bonds, the AA-stacking can be further divided into AA parallel (AA-P) when A1-A2 is parallel to A3-A4 and AA non-parallel (AA-NP) when A1-A2 and A3-A4 are crossed with respect to each other (Fig. 1(a)). For AB-stacking, the four silicon atoms are non-coplanar. The AB-stacking has three forms depending on the orientation of the bonds A1-A2 and A3-A4: AB-parallel (AB-P), AB non-parallel (AB-NP), and AB-hybrid (Fig. 1(b)). The coordinates of the four silicon atoms for the five local minima are listed in table 1. It is worth mentioning that the AA-stacking exhibits a slightly lower total energy than the AB-stacking, indicating a more stable structure. Furthermore, our calculations reveal that the induced biaxial strain can only open an energy band gap in AA-stacking, thus the remaining discussions are focused on AA-stacking.

In both AA-P and AA-NP structures, the vertical distances along the z-axis (Fig. 1(a)) between A1 and A2, and A3 and A4 are equal (Table 1). Without losing generality, the buckling height is measured by the vertical distance between atoms A1 and A2. The biaxial strain/stress is applied along the unit vectors with their directions shown in Fig. 2 (inset). Figure 2 shows the buckling heights of the AA-P and AA-NP versus the applied biaxial strain. In the calculations, the strain ($\varepsilon$) is induced by changing the lattice constant (*a*) of the unit cell from the relaxed lattice constant ($a_0 = 3.866$ Å) and is equal to $\varepsilon = \frac{a - a_0}{a_0}$. An increase of the biaxial strain leads to a reduction of the buckling height. As the strain increases to 5.17%, a sudden drop of the buckling height is observed in AA-P. Upon further increase of the strain, the buckling height reaches and maintains at zero. Similar drop of the buckling height has been observed for AA-NP when the strain is about 7.76%. After the buckling height reaches zero, both the AA-P and AA-NP structures become a flat 2D structure with hexagonal symmetry similar to their carbon counterpart graphene. It should be pointed out that both the single atomic layered silicene and the BiSF with AB-stacking remain in the buckled structure under biaxial strain/stress.

Energy band diagrams of AA-P and AA-NP are shown in figure 3 as a function of the in-plane strain. Significant changes have been observed as the biaxial strain reaches 5.17% for AA-P (Fig. 3(b)) and 7.76% for AA-NP (Fig. 3(g)); this is evident by showing a single conduction band at M-point and a single valence band between K and Γ points near the Fermi energy level. These values equal to the critical strains for zero buckling height for AA-P and AA-NP (5.17% for AA-P and 7.76% for AA-NP) in figure 2. This coincidence indicates a strong correlation between the buckling height and the energy band gap. Since the AA-P and AA-NP structures become the same as the strain is above the critical values, they have the same band diagrams as shown in figures 3 (c) and (g), and (d) and (h). The bottom of the conduction band shifts upwards to a higher energy while the valence band moves downwards to a lower energy as the biaxial strain becomes larger. However, due to the crossing at the Fermi energy level of both the conductance and valence bands, the BiSF is a semimetal as the strain is less than 10.35%.

Further increase of the strain (Fig. 4) leads to an energy band gap opening with a value of 18.0 meV when $\varepsilon = 11.2\%$ (Fig. 4 (a)), and reaches a maximum of 168.0 meV when $\varepsilon = 14.3\%$ (Fig. 4 (b)). However, due to the lowering of the band $L_1$ and up-moving of the bands $L_2$ and $L_3$, the band gap opening is reduced for $\varepsilon = 14.8\%$ (Fig. 4 (c)), and eventually disappears for $\varepsilon = 16.4\%$ (Fig. 4 (d)). The opened energy band gap shows a linear dependency on the strain (Fig. 5): increasing for $11.2\% \leq \varepsilon \leq 14.3\%$, and decreasing for $14.3\% \leq \varepsilon \leq 16.4\%$ (Fig. 5).

The transmission spectrum and current-voltage (I-V) curves at the maximum band gap opening are shown in figure 5. The band gap opening is evident and manifested by the zero transmission efficiency close to the Fermi energy level. This is in a perfect agreement with its energy band diagram (Fig. 4(b)). Due to the existence of the energy band gap, the I-V curve exhibits a plateau of zero current intensity in the vicinity of zero applied voltage.

**Conclusions**

In this work, the structural and electronic transport properties of BiSF were investigated. Five energetically favorable structures of BiSF with AA- and AB- stacking configurations have been obtained. The bucking height of AA-stacked BiSF decreases to zero as the applied in-plane strain exceeds 5.17% for AA-P and 7.76% for AA-NP structures. Upon further increase of the strain, AA-P and AA-NP converge to a single structure and start to have an energy band gap. A correlation has been observed between the strain-induced buckling height reduction and the band gap appearing. It turns out that the range of strain to open the band gap is 10.7% ~ 15.4%, and the maximum energy band gap opening is about 168.0 meV. Hence, we propose the BiSF structure as a candidate for having a band gap; the biaxial strain could be induced by growth on the appropriate substrate. Preliminary calculations indicate the band gap would be preserved.

Figure captions:

Figure 1 (a) Side views of an unit cell of AA-P and AA-NP with unit vectors "A", "B", and "C"., and the top view of a 3 by 3 supercell of AA-P and AA-NP with lattice constant a=3.866 Å. The top views of AA-P and AA-NP are identical. (b) Side views and top view of AB-P, AB-NP, and AA-NP. A1, A2, A3 and A4 are the four silicon atoms with their coordinates listed in table I.

Figure 2 Buckling height of AA-P (triangle) and AA-NP (black circle) with induced bi-axial strain. The inset shows the direction of the applied bi-axial strain/stress.

Figure 3 Band diagrams of AA-P (a), (b) (c) and (d), and of AA-NP (e), (f), (g), and (h) with various induced tensile strain. The band diagrams of flat AA-P ((b), (c) and (d)) and flat AA-NP ((g) and (f) are significantly different from the bucked structures, e.g. (a) for AA-P and (e) and (f) for AA-NP. As buckling height equals to zero, the band diagrams of AA-P are identical to those of AA-NP, manifested by the indistinguishable diagrams between (c) and (g), (d) and (h).

Figure 4 Band gap opening versus various tensile strains. (a) Band gap is obtained by considering C1 and V1 bands when tensile strain is 11.2%. (b) Band gap reaches its maximum when tensile strain is 14.3%. (c) Band gap starts to decrease due to the lowering of the $L_1$ band and is counted between the bottom of $L_1$ and the top of $V_1$. (d) Band gap disappears due to the overlapping of $L_1$ and $L_2$ when tensile strain is above 16.4%.

Figure 5 Band gap opening of AA-stacking by applying various bi-axial strains. The insets show the I-V curve (top-left) and the transmission spectrum (bottom-right) of BiSF when the energy band gap is opened to maximum.

Figure 1

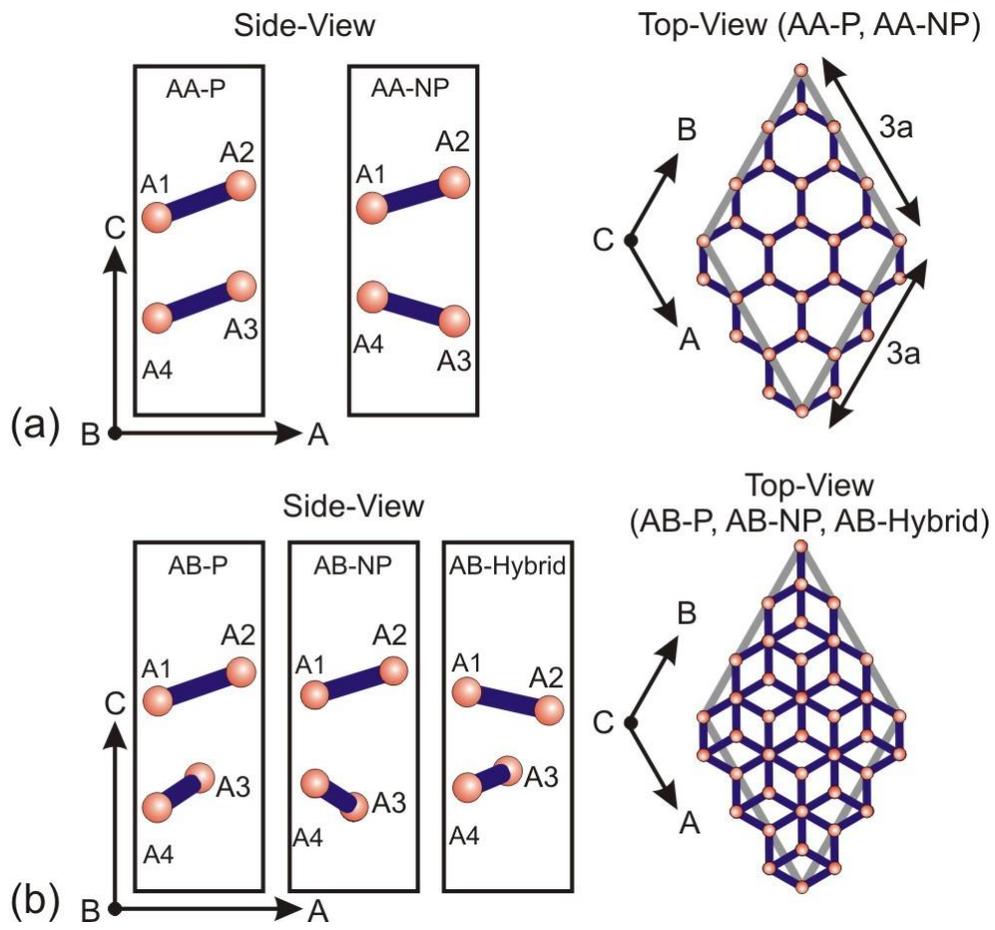

Figure 2

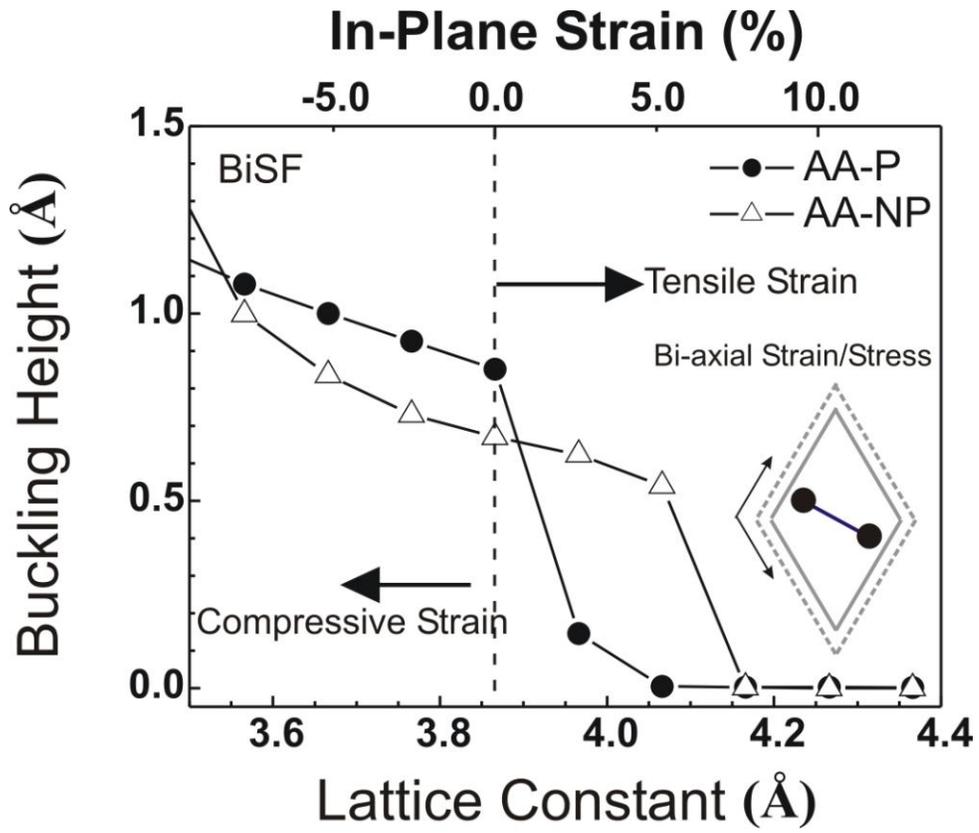

Figure 3

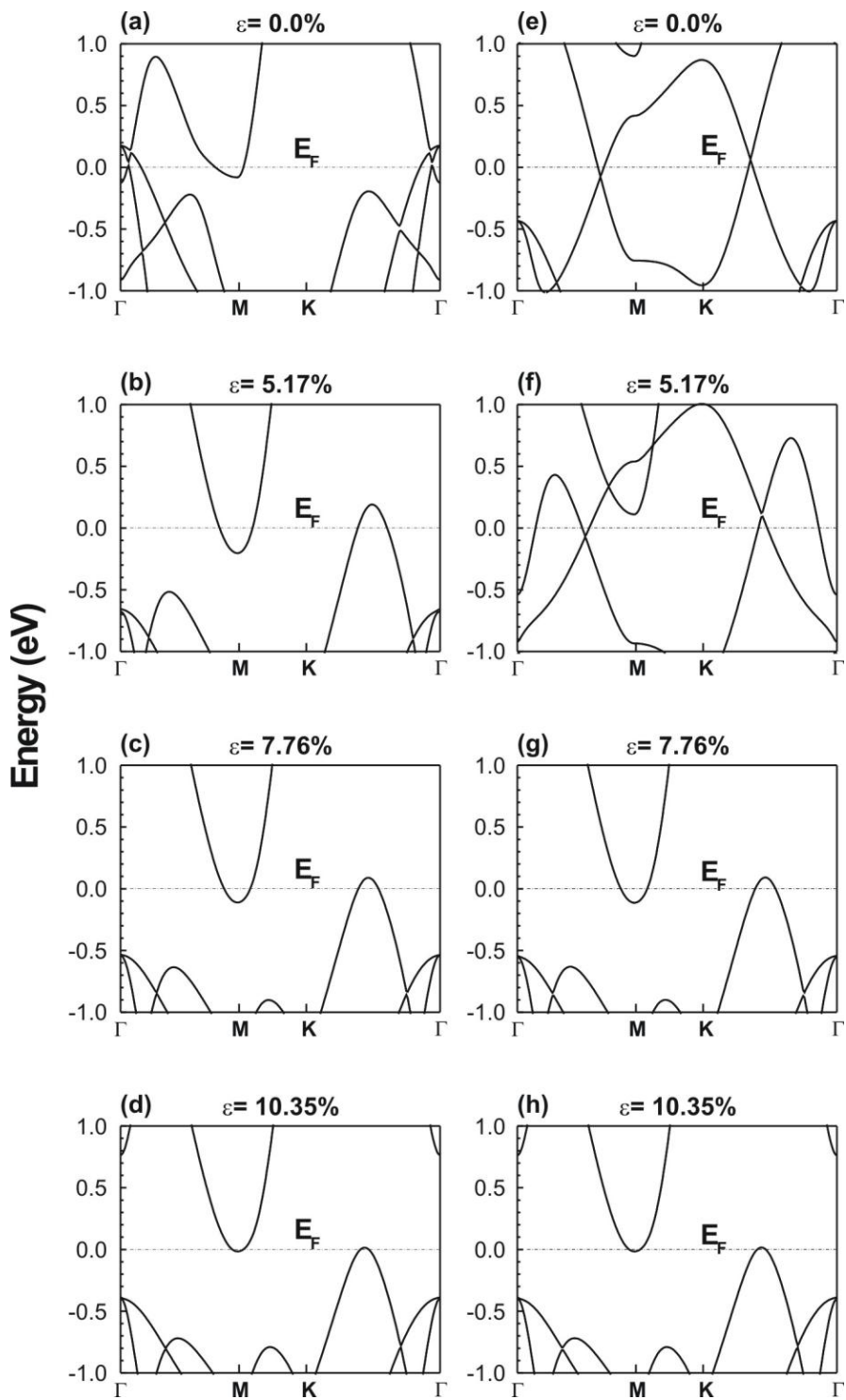

Figure 4

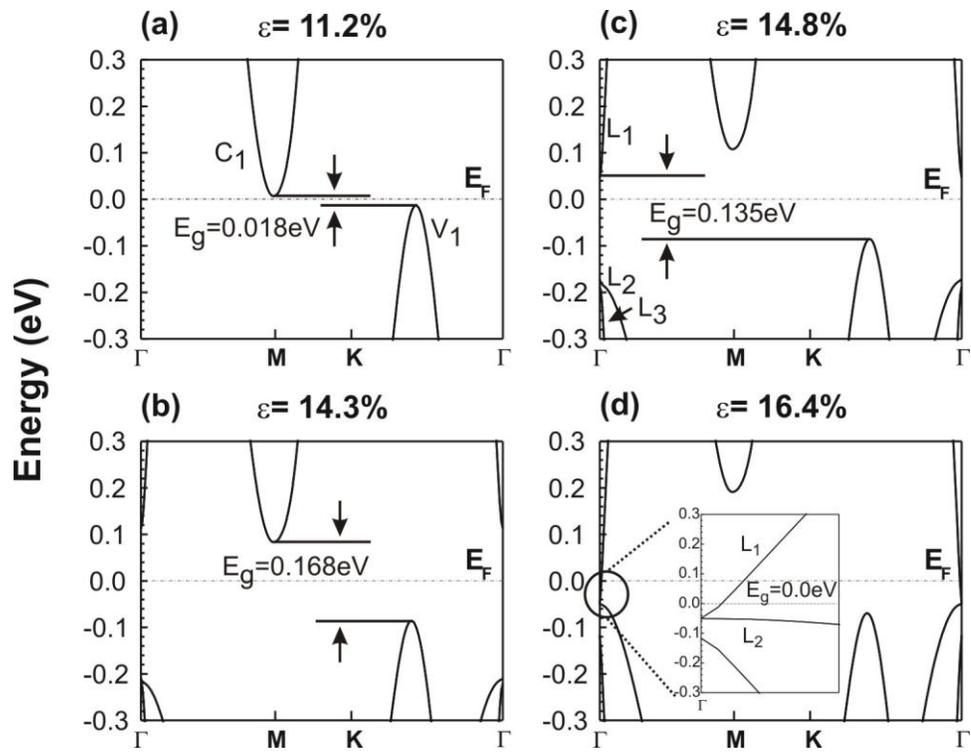

Figure 5

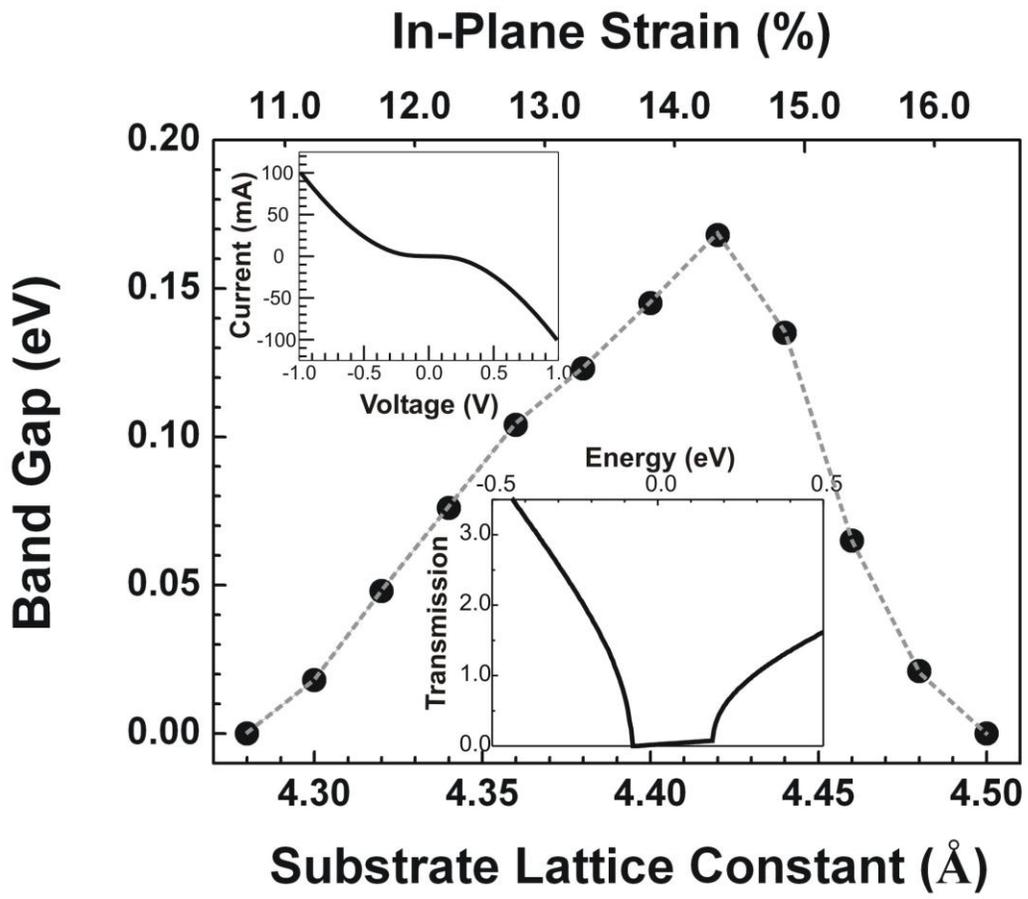

Table I. Coordinates of the three silicon atoms in BiSF's unit cell with the forth atom sitting at the origin. The A4 sits at the origin of the Cartesian coordinate.

|  | A1(Å) | | | A2(Å) | | | A3(Å) | | |
| --- | --- | --- | --- | --- | --- | --- | --- | --- | --- |
|  | x | y | z | x | y | z | x | y | z |
| AA-P | 0.0 | 0.0 | 2.74 | 1.93 | -1.13 | 3.59 | 1.93 | -1.11 | 0.85 |
| AA-NP | 0.0 | 0.0 | 2.45 | 1.93 | -1.12 | 3.12 | 1.93 | -1.12 | -0.67 |
| AB-P | 0.0 | 0.0 | 2.87 | 1.94 | -1.14 | 3.65 | 1.94 | 1.12 | 0.71 |
| AB-NP | 0.0 | 0.0 | 2.50 | 1.93 | -1.13 | 3.18 | 1.93 | 1.12 | -0.68 |
| AB-Hybrid | 0.0 | 0.0 | 2.61 | 1.93 | -1.13 | 2.10 | 1.93 | 1.11 | 0.53 |